\documentclass[12pt]{article}
\usepackage{fullpage}
\usepackage{amsmath}
\usepackage{amssymb}
\usepackage[dvips]{epsfig}

\def\##1{\underline{#1}}
\def\=#1{\underline{\underline{#1}}}

\def\+#1{\underline{\bf #1}}
\def\*#1{\underline{\underline{\bf #1}}}

\def\r#1{(\ref{#1})}
\def\l#1{\label{#1}}
\def\c#1{\cite{#1}}

\def\le{\left(}
\def\ri{\right)}
\def\les{\left[}
\def\ris{\right]}
\def\lec{\left\{}
\def\ric{\right\}}

\def\.{\mbox{ \tiny{$^\bullet$} }}

\def\epso{\epsilon_{\scriptscriptstyle 0}}

\def\muo{\mu_{\scriptscriptstyle 0}}

\def\curl{\nabla\times}

\def\curl{\nabla\times}

\def\gmet{g_{\alpha\beta}}

\def\tr{(ct,\#r)}
\def\ok{(\omega/c,\#k)}

\def\mrbh{m_{\mbox{\tiny{rbh}}}}
\def\arbh{a_{\mbox{\tiny{rbh}}}}

\begin{document}

\begin{center}

{\bf {\Large  Electromagnetic Negative--Phase--Velocity Propagation in  the
Ergosphere of a Rotating Black Hole }}

 \vspace{10mm} \large

Tom G. Mackay\footnote{Corresponding Author. Fax: + 44 131 650
6553; e--mail: T.Mackay@ed.ac.uk.} \\
{\em School of Mathematics,
University of Edinburgh, Edinburgh EH9 3JZ, UK}\\
\bigskip
 Akhlesh  Lakhtakia\footnote{Corresponding Author. Fax: +1 814 863
4319; e--mail: akhlesh@psu.edu; also
 affiliated with Department of Physics, Imperial College, London SW7 2 BZ,
UK}\\
 {\em Department of Engineering Science and
Mechanics\\ Pennsylvania State University, University Park, PA
16802--6812, USA}\\
\bigskip
Sandi
Setiawan\footnote{Fax: + 44 131
650 6553; e--mail: S.Setiawan@ed.ac.uk.}\\
{\em School of Mathematics,
University of Edinburgh, Edinburgh EH9 3JZ, UK}\\

\end{center}

\vspace{4mm}

\normalsize

\begin{abstract}
\noindent We report on the occurrence of negative--phase velocity
(NPV) planewave propagation in the ergosphere of a rotating black
hole. By implementing the Kerr metric, it is demonstrated that
regions of NPV propagation are concentrated at the equator of the
ergosphere, while NPV propagation is less common towards the polar
regions. Increasing the  angular velocity of the black hole
exaggerates the NPV concentration at the equator. NPV propagation
is not observed outside the stationary limit surface.

\end{abstract}

\noindent {\bf Keywords:}  General theory of relativity, Negative phase
velocity, Poynting vector, Kerr spacetime

\section{Introduction}

The term \emph{negative--phase velocity} (NPV) refers to a mode of
electromagnetic planewave propagation in which the phase velocity
vector is directed opposite to the direction of time--averaged
energy flux \c{LMW02,LMW03}. Of the many interesting consequences of NPV
propagation, it is the phenomenon of negative refraction which has
recently attracted much attention  in the electromagnetics/optics
and materials research communities \c{Pendry04}. This interest may
be traced back to the first experimental reports of negative
refraction in 2001, involving the microwave illumination of an
artificial \emph{metamaterial} \c{SSS}.

We have recently begun to examine the possibility of
gravitationally--assisted NPV propagation in vacuum \c{LM04,LMS05}. In
the absence of a `material', the source of the NPV may be provided
by the curvature of spacetime. The electrodynamic formulation adopted
is founded on the formal analogy  between
 (a) electromagnetic  propagation in curved spacetime in free space; and (b)
electromagnetic  propagation in flat spacetime in a fictitious,
instantaneously responding
 bianisotropic medium.
This approach~---~which was originally put forward by Tamm and
subsequently described by numerous authors
\c{LL}--\c{Burko}~---~enables wave propagation in curved spacetime
to be examined using standard electromagnetic techiques. Thereby,
the feasibility of NPV propagation for a general gravitomagnetic
metric has been established \c{MLS05}.

 We further develop the concept of gravitationally--assisted
 NPV propagation in this paper with a detailed numerico--theoretical
 study based  on
  a particular gravitomagnetic metric, namely the Kerr metric \c{Kerr}.
Specifically, we  explore the spacetime surrounding a rotating
black hole, in search of regions where NPV propagation occurs. Our
search is restricted to the physically probeable spacetime outside
the outer event horizon. The theoretical framework is described in
Section 2, numerical results are presented and analyzed in Section
3, and conclusions are itemized and discussed in Section 4.

\section{Theoretical framework}

The electrodynamics of the classical vacuum in a curved spacetime
provides the setting for our numerico--theoretical study. Thus, we consider the
covariant and the contravariant electromagnetic field tensors
$f_{\alpha\beta}$ and $h^{\alpha\beta}$, respectively, which
appear in the source--free covariant Maxwell equations\footnote{Greek indexes
take the values 0, 1, 2 and 3; Roman indexes take the values 1, 2
and 3; $x^0 = ct$ where $c$ is the speed of light in vacuum in the
absence of all gravitational fields; whereas $x^{1,2,3}$ are the
three spatial coordinates.}
\begin{equation}
\label{ME1} \left.
\begin{array}{l}
 f_{\alpha\beta;\nu} +
f_{\beta\nu;\alpha}+f_{\nu\alpha;\beta} = 0\, \\ \vspace{-8pt} \\
h^{\alpha\beta}_{\quad;\beta} = 0\,
\end{array}
\right\} ,
\end{equation}
where the subscript $_{;\nu}$ indicates the covariant derivative
with respect to the $\nu$th spacetime coordinate $x^\nu$. The
spacetime curvature is described by the  metric $\gmet$ with
signature $\le +,-,-,- \ri$.

\subsection{Electrodynamic formulation}

We follow the standard approach \c{Skrotskii}--\c{Mashhoon}~---~originally proposed by
Tamm~---~wherein the covariant equations \r{ME1} are replaced by
their noncovariant equivalents
\begin{equation}
\label{ME1_noncov} \left. \begin{array}{l} f_{\alpha\beta,\nu} +
f_{\beta\nu,\alpha}+f_{\nu\alpha,\beta} = 0\, \\ \vspace{-6pt} \\
\les \le -g \ri^{1/2}  h^{\alpha\beta} \ris_{,\beta} = 0\,
\end{array}
\right\} ,
\end{equation}
with $g = \mbox{det} \les g_{\alpha \beta} \ris $ and the
subscript $_{,\nu}$ denoting ordinary differentiation with respect
to the $\nu$th spacetime coordinate. More conveniently, we express
\r{ME1_noncov} in the familiar form
\begin{equation}
\label{ME2} \left.
\begin{array}{cc}
B_{\ell,\ell} = 0\,, & B_{\ell,0} + \varepsilon_{\ell mn} E_{m,n}
= 0\\ \vspace{-8pt} \\ D_{\ell,\ell} = 0\,, & -D_{\ell,0} +
\varepsilon_{\ell mn} H_{m,n} = 0
\end{array}\right\}\,,
\end{equation}
where $\varepsilon_{\ell mn}$ is the three--dimensional
Levi--Civita symbol. The conventional electromagnetic field
vectors $\#E$ , $\#B$ , $\#D$ and $\#H$ are the 3
vector--equivalents associated with the respective
  components
\begin{equation}
\left.
\begin{array}{ll}
E_\ell = f_{\ell 0}\,, \\ B_\ell = (1/2) \varepsilon_{\ell mn}f_{mn}\\
D_\ell= \le -g \ri^{1/2} h^{\ell 0}\,, \\  H_\ell=(1/2)
\varepsilon_{\ell mn} \le -g \ri^{1/2} h^{mn}
\end{array}\right\}\,
\label{bbb}
\end{equation}
 introduced in \r{ME2}. Thus, in \r{ME2}, $B_{\ell,\ell}$ and
$D_{\ell,\ell}$ represent the divergences of $\#B$ and $\#D$,
whereas the $\ell$th vector component of the curls  of $\#E$ and
$\#H$ are represented as $\varepsilon_{\ell mn} E_{m,n}$ and
$\varepsilon_{\ell mn} H_{m,n}$, with the derivatives with respect
to time of the $\ell$th vector component of $\#B$ and $\#D$ being
represented as $B_{\ell,0}$ and $D_{\ell,0}$. The electromagnetic
field vector components
 satisfy the constitutive relations
\begin{equation}
\label{CR2} \left.
\begin{array}{l}
D_\ell = \gamma_{\ell m} E_m + \varepsilon_{\ell mn}\,\Gamma_m\,H_n\\[6pt]
B_\ell =    \gamma_{\ell m} H_m - \varepsilon_{\ell mn}\,
\Gamma_m\,  E_n
\end{array}\right\}\,,
\end{equation}
where the tensor $\gamma_{\ell m}$ components and vector
$\Gamma_m$ components are defined as
\begin{equation}
\label{akh1} \left.\begin{array}{l} \gamma_{\ell m}
= \displaystyle{- \le -{g} \ri^{1/2} \, \frac{{g}^{\ell m}}{{g}_{00}}}\\[6pt]
\Gamma_m= \displaystyle{\frac{g_{0m}}{g_{00}}}
\end{array}\ric
\,.
\end{equation}
Thus, in terms of the noncovariant formulation \r{ME1_noncov},
vacuum is represented by the fictitious,
instantaneously--responding, bianisotropic medium characterized by
\r{CR2}.

For later use, we recast \r{ME2} and \r{CR2} in  3 $\times$ 3
dyadic/ 3 vector form  as
\begin{equation}
\label{eq1} \left.
\begin{array}{l}
\curl \#E\tr + \displaystyle{\frac{\partial}{\partial t}} \#B\tr =
0\,\\ \vspace{-8pt} \\ \curl\#H\tr -
\displaystyle{\frac{\partial}{\partial t}} \#D\tr = 0\,
\end{array}
\right\}
\end{equation}
and
\begin{equation}
\left.
\begin{array}{l}
 \label{eq2}
\#D\tr = \epso\,\=\gamma\tr\. \#E\tr - \displaystyle{\frac{1}{c}}\, \#\Gamma\tr\times \#H\tr\,\\
\vspace{-8pt} \\ \#B\tr = \muo\,\=\gamma\tr\. \#H\tr +
\displaystyle{\frac{1}{c}}\,\#\Gamma\tr\times \#E\tr\,
\end{array}
\right\},
\end{equation}
respectively, wherein $\=\gamma\tr$ is the dyadic--equivalent of
$\gamma_{\ell m}$,
 $ \#\Gamma\tr$ is the vector--equivalent of $\Gamma_m$;
the scalar constants $\epso$ and $\muo$ denote the permittivity
and permeability of vacuum in the absence of a gravitational
field; and SI
 units are adopted.
 In moving from  \r{ME1} to
\r{eq1} and \r{eq2},
 the spacetime coordinate $x^\alpha$ has been separated into
   space $\#r$ and time $t$ coordinates.
The  vector cross products and dyadic--vector dot products in
\r{eq1} and \r{eq2}
 apply to vectors and dyadics associated with the three spatial dimensions of flat spacetime \c{Chen}.
    However, it is emphasized that
 the spacetime underlying
\r{eq1} and \r{eq2} remains curved.

\subsection{Kerr spacetime}

We consider  a specific gravitomagnetic  spacetime, namely the
Kerr spacetime \c{Kerr}, which  describes a rotating black hole.
 In Cartesian
coordinates~---~$\#r\equiv (x,y,z)$~---~the components of the Kerr metric are given as \c{Inverno}
\begin{eqnarray}
g_{00} &=& 1 - \Delta \,, \l{g00} \\
g_{01} &=& - \Delta \, \frac{R x + \arbh y}{R^2 + \arbh^2}\,,\\
g_{02} &=& - \Delta \, \frac{Ry - \arbh x}{R^2 + \arbh^2} \,,\\
g_{03} &=& -\Delta \, \frac{z}{R} \,,\\
g_{11} &=& -1 - \Delta \le \frac{R x + \arbh y}{R^2 + \arbh^2} \ri^2\,,\\
g_{12} &=& - \Delta  \frac{ \le R x + \arbh y \ri \le Ry - \arbh x \ri }{ \le R^2 + \arbh^2 \ri^2} \,,\\
g_{13} &=& - \Delta  \frac{ \le R x + \arbh y \ri z }{ \le R^2 + \arbh^2 \ri R} \,,\\
g_{22} &=& -1 - \Delta \le \frac{\arbh x - R y}{R^2 + \arbh^2} \ri^2\,,\\
g_{23} &=& - \Delta  \frac{ \le R y - \arbh x \ri z }{ \le R^2 + \arbh^2 \ri R} \,,\\
g_{33} &=& -1 - \Delta \le \frac{z}{R} \ri^2\,, \l{g33}
\end{eqnarray}
with  $g_{\alpha \beta} = g_{\beta \alpha}$.  The quantity $R$ is
given implicitly via
\begin{equation}
R^2 = x^2 + y^2 + z^2 - \arbh^2 \les 1 - \le \frac{z}{R} \ri^2
\ris\,;
\end{equation}
also,
\begin{equation} \Delta = \frac{2 \mrbh  R^3}{R^4 + \le \arbh z \ri^2}\,.
\end{equation}
The components \r{g00}--\r{g33} of the metric describe a black
hole  of geometric mass $\mrbh$, rotating about the Cartesian $z$
axis, in conventional units. The term $\arbh$ is a measure of the
black hole's angular velocity (sometimes referred to as the specific angular momentum).

We note that the Kerr metric is often represented in terms of
Boyer--Lindquist coordinates \cite[Sec. 19.3]{Inverno}. However,
for the planewave analysis undertaken in the following sections,
the Cartesian representation \r{g00}--\r{g33} is more convenient.

Using the definitions \r{akh1} together with \r{g00}--\r{g33}, we obtain
\begin{eqnarray}
\l{gamma_11} \gamma_{11} &=& \delta \les R^4 + \le \arbh z \ri^2
\ris \les 1 -
\Delta \le \frac{R x + \arbh y}{R^2 + \arbh^2} \ri^2 \ris \,,\\
\gamma_{12} &=& - 2\delta \, \mrbh R^3  \frac{\le R x + \arbh y \ri
\le R y -
\arbh x \ri }{\le R^2 + \arbh^2 \ri^2}\,\,\\
\gamma_{13} &=& - 2\delta \, \mrbh R^2 z  \frac{ R x + \arbh y
 }{ R^2 + \arbh^2 }\,\,\\
\gamma_{22} &=& \delta \les R^4 + \le \arbh z \ri^2 \ris \les 1 -
\Delta \le \frac{\arbh x - R y}{R^2 + \arbh^2} \ri^2 \ris \,,\\
\gamma_{23} &=& -2 \delta \, \mrbh R^2 z  \frac{ R y - \arbh x
 }{ R^2 + \arbh^2 }\,\,\\
\gamma_{33} &=& \delta \les R^4 + \le \arbh z \ri^2 \ris \les 1 -
\Delta \le \frac{z}{R } \ri^2 \ris \,, \l{gamma_33}
\end{eqnarray}
with $\gamma_{\ell m} = \gamma_{m \ell}$, and
\begin{eqnarray}
\l{Gamma_1}
\Gamma_1 &=& - 2\delta\,  \mrbh R^3 \, \le \, \frac{ R x + \arbh y}{R^2 + \arbh^2 } \, \ri,\\
\Gamma_2 &=& -2 \delta \, \mrbh R^3 \, \le \, \frac{ R y - \arbh x}{R^2 + \arbh^2 } \, \ri,\\
\Gamma_3 &=& - 2\delta \, \mrbh R^2 z\,, \l{Gamma_3}
\end{eqnarray}
wherein
\begin{equation}
\delta = \frac{1}{R^4 + \le \arbh z \ri^2 - 2\mrbh R^3}\,.
\end{equation}

We concentrate on the regime $\mrbh^2 > \arbh^2$. Furthermore,  as
our interest lies in physically observable phenomenons, we
restrict attention to the region $R
> R_+$ where
\begin{equation}
R_+ = \mrbh + \sqrt{ \mrbh^2 - \arbh^2}
\end{equation}
represents the outer event horizon of the black hole. Finally, we
note that $R_+ \leq R_{S_+}$ where
\begin{equation}
R_{S_+}  = \mrbh + \sqrt{ \mrbh^2 - \le \, \frac{\arbh z}{R_{S_+}}
\,\ri^2}
\end{equation}
delineates the stationary limit surface of the black hole.

\subsection{Planewave propagation}

Let us consider the propagation of plane waves  in the
medium characterized by the constitutive relations \r{eq2}. Our
strategy is to focus  upon an arbitrary  neighbourhood $\cal R$ whose location
 is  specified by the
 Cartesian coordinates $\le \tilde{x},
\tilde{y}, \tilde{z} \ri$. The neighbourhood is presumed to be
sufficiently small that the nonuniform metric $g_{\alpha \beta}$
may  be approximated by the uniform metric ${\tilde g}_{\alpha
\beta}$ throughout  $\cal R$. Thus, within $\cal R$, the uniform
3$\times$3 dyadic $
 \={\tilde\gamma} \equiv \left. \=\gamma \,\right|_{\cal R} $
plays the role of the metric  with components
\r{gamma_11}--\r{gamma_33}, whereas the uniform 3 vector $
\#{\tilde\Gamma} \equiv \left. \#\Gamma \, \right|_{\cal R} $
plays the role of the vector with components
\r{Gamma_1}--\r{Gamma_3}. As a consquence of the
uniform metric approximation, henceforth we restrict our consideration
to  electromagnetic wavelengths  which are small
compared with the linear dimensions of the neighbourhood $\cal R$.

Within the neighbourhood $\cal R$, the electromagnetic fields are
 decomposed via the
  three--dimensional Fourier transforms
\begin{equation}
\left.
\begin{array}{l}
\#E (ct, \#r) = \displaystyle{ \frac{1}{c} \int_{-\infty}^\infty
\int_{-\infty}^\infty \int_{-\infty}^\infty
 \#{\sf E}(\omega/c, \#k) \exp\les i(\#k \. \#r - \omega t)\ris \, d\omega \, dk_1  \, dk_2\,} \\ \vspace{-2mm} \\
\#H (ct, \#r)  = \displaystyle{ \frac{1}{c} \int_{-\infty}^\infty
\int_{-\infty}^\infty  \int_{-\infty}^\infty
 \#{\sf H} (\omega/c, \#k) \exp\les i(\#k \. \#r - \omega t )\ris \, d \omega \, dk_1 \, dk_2\, }
\end{array}
\right\}, \label{der}
  \end{equation}
 where the wavevector
$\#k$ is the Fourier variable corresponding to $\#r$,
  $i=\sqrt{-1}$ and $\omega$ is the
 usual temporal frequency. The wavevector component
  $k_3$ is
  determined by
combining \r{der} with the Maxwell curl postulates \r{eq1} and
then solving the resulting 4 $\times$ 4 eigenvalue problem
\c{Lopt92}. We take $k_3 \in \mathbb{R}$ and thereby consider only
propagating (non--evanescent) plane waves.

The associated rate of energy flow is calculated from the
complex--valued phasors of the electric and magnetic fields,
 $\#{\sf E} (\omega/c, \#k)$ and  $\#{\sf H} (\omega/c, \#k)$
 respectively,
by means of the time--averaged Poynting vector\footnote{$\mbox{Re}
\lec \cdot \ric$ denotes the real part while the asterisk denotes
the complex conjugate.}
\begin{equation}
 \langle \, \#{\sf P} (\omega/c, \#k) \, \rangle_t =  \frac{1}{2} \, \mbox{Re}
 \, \lec \, \#{\sf E} (\omega/c, \#k) \times \#{\sf H}^* (\omega/c,
 \#k)\,\ric.
\end{equation}
If the  phase velocity vector casts a negative projection onto
the time--averaged Poynting vector then we have so--called
negative phase velocity (NPV); i.e., NPV is signalled by
\begin{equation}
\#k \. \langle \, \#{\sf P} (\omega/c, \#k) \, \rangle_t < 0\,.
\end{equation}

The key steps involved in deriving a general expression for  the
quantity $\#k \. \langle \, \#{\sf P} (\omega/c, \#k) \,
\rangle_t$  follow the corresponding steps for a simply moving
medium which is isotropic dielectric--magnetic at rest \c{Chen},
\c{ML04a}.
 We begin by combining the Fourier decompositions \r{der} with
the constitutive relations \r{eq2} and the Maxwell curl postulates
\r{eq1}. Thereby, we find
\begin{eqnarray}
\#p \times \#{\sf E}\ok &=& \omega\muo \,\={\tilde\gamma} \.\#{\sf
H}\ok\,, \l{h12}
\\
\#p \times \#{\sf H}\ok &=& -\omega\epso \,\={\tilde\gamma} \.
\#{\sf E}\ok\,\l{e12}
\end{eqnarray}
where the vector
\begin{equation}
\#p = \#k -  \frac{\omega}{c} \,\#{\tilde\Gamma}\,
\end{equation}
has been introduced.  Upon substituting \r{h12} into \r{e12} and carrying out some
algebraic manipulations, we obtain
  the eigenvector equation
\begin{equation}
\lec \les \le\frac{\omega}{c}\ri^2 \, \vert \={\tilde\gamma}\vert
- \#p\. \={\tilde\gamma}\.\#p\ris\=I +\#p\,\#p\.
\={\tilde\gamma}\ric \. \#{\sf E}\ok  = \#0\,
 \label{eee2}
 \end{equation}
and the corresponding dispersion relation
\begin{equation}
\label{dispeq} \les\#p \. \,\,\={\tilde\gamma} \. \#p -
\le\frac{\omega}{c}\ri^2 \, \vert \,\={\tilde\gamma} \,
\vert\ris^2 = 0\,.
\end{equation}
Here, $\=I$ is the identity
 dyadic, and
$ \vert \, \={\tilde\gamma}\,\vert $ denotes the determinant of $
\={\tilde\gamma}$.

To solve \r{eee2}, we exploit the orthogonality condition
 \begin{equation}
\#p\.\={\tilde\gamma}\. \#{\sf E}\ok = 0\,. \label{eee3}
\end{equation}
which follows from \r{dispeq}. The general solution to the
electric phasor equation \r{eee2} is expressible as the sum
\begin{equation}
\l{e_gs} \#{\sf E}\ok =
 A_a\ok\,
 \#{\sf e}_a\ok\,
+
 A_b\ok\,
 \#{\sf e}_b\ok\,
\end{equation}
where the unit vectors
\begin{equation}
\left.
\begin{array}{l}
 \#{\sf e}_a
= \displaystyle{\frac{\={\tilde\gamma}^{-1} \. \#w}
{\vert\={\tilde\gamma}^{-1} \. \#w\vert}} \\ \\\  \#{\sf e}_b =
\displaystyle{\frac{\={\tilde\gamma}^{-1}\. \le \#p \times \#{\sf
e}_a \ri}{\vert\={\tilde\gamma}^{-1}\. \le \#p \times \#{\sf e}_a
\ri\vert}}
\end{array}
\right\}
 \,. \l{e_12}
\end{equation}
are chosen in accordance with \r{eee3}. The  unit vector $\#w$
is orthogonal to $ \#p$, i.e., $\#w\.\#p=0$,  but is otherwise arbitrary.
For calculating the numerical results presented in the
following section, we chose
\begin{equation}
\#w = \left\{ \begin{array}{lcr} \displaystyle{\frac{1}{\sqrt{2
p^2_z + \le p_x + p_y \ri^2} } }\le p_z, p_z, -p_x-p_y \ri &
\mbox{for} & p_z \neq 0
\\ & \\
\displaystyle{ \frac{1}{\sqrt{ p_y^2 + p^2_x}}} \le p_y, -p_x, 0
\ri & \mbox{for} & p_z = 0
\end{array}
\right.
,
\end{equation}
with $\#p\equiv \le p_x, p_y, p_z \ri$ in the Cartesian coordinate
system. Initial and boundary conditions fix the
complex--valued amplitudes
 $ A_{a,b} \ok$.
The  general solution to the magnetic phasor equation
corresponding to \r{eee2} follows directly from \r{e_gs} upon
substitution into \r{h12}.

Finally, combining \r{eee2} and \r{h12} yields the central
result
\begin{eqnarray}
\#k \. \langle \#{\sf P}\rangle_t &=& \frac{1}{2 \omega \muo \vert
\={\tilde\gamma} \vert} \le \vert A_a \vert^2 \#{\sf e}_a \.
\={\tilde\gamma} \. \#{\sf e}_a +  \vert A_b \vert^2 \#{\sf e}_b
\. \={\tilde\gamma} \. \#{\sf e}_b \ri \, \#k \. \={\tilde\gamma}
\. \#p \,. \label{ppp1}
\end{eqnarray}
For the Kerr metric, we observe that
\begin{equation}
\vert \, \=\gamma \, \vert = \delta^2 \les R^4 + \le \arbh z \ri^2
\ris^2\,.
\end{equation}
Therefore, it follows that  $\vert \, \={\tilde\gamma}\, \vert >
0$ and the sufficient condition
\begin{equation} \l{NPV_cond}
\left.
\begin{array}{l}
P_a < 0 \\
P_b < 0
\end{array}
\right\}
\end{equation}
 for NPV propagation arises,
wherein
\begin{equation}
\left. \l{PaPb}
\begin{array}{l}
P_a = \le \, \#{\sf e}_a \. \={\tilde\gamma} \. \#{\sf e}_a \, \ri
\le \, \#k \. \={\tilde\gamma} \. \#p \, \ri \\ \vspace{-6pt}
\\ P_b = \le \, \#{\sf e}_b \. \={\tilde\gamma} \. \#{\sf e}_b \,
\ri \le \, \#k \. \={\tilde\gamma} \. \#p \, \ri
\end{array}
\right\}.
\end{equation}

The dispersion relation \r{dispeq} yields  two wavenumbers $k =
k^\pm$ for the arbitrarily oriented wavevector $\#k = k \hat{\#k}$
with $\hat{\#k} = \le \sin \theta \cos \phi, \sin \theta \sin
\phi, \cos \theta \ri$. These are
\begin{equation} \l{wave_nos}
k^\pm = \frac{\omega}{c} \le \frac{\hat{\#k} \. \={\tilde\gamma}
\. \#{\tilde \Gamma} \pm \sqrt{ \le \, \hat{\#k} \.
\={\tilde\gamma} \. \#{\tilde \Gamma} \, \ri^2 - \hat{\#k} \.
\={\tilde\gamma} \. \hat{\#k} \le \#{\tilde\Gamma} \.
\={\tilde\gamma} \. \#{\tilde\Gamma} - \vert \,\={\tilde\gamma} \,
\vert\, \ri}}{\hat{\#k} \. \={\tilde\gamma} \. \hat{\#k}}\ri\,.
\end{equation}
As only propagating (non--evanescent) plane waves are considered here, we have $k^\pm \in \mathbb{R}$.
In order to establish whether $\#k \. \langle \#{\sf P}\rangle_t <
0$ for the wavenumbers $k^\pm$,  we resort to numerical
evaluations of \r{ppp1} and \r{wave_nos}.

\section{Numerical results}

The prospects for NPV propagation are investigated within the
gravitational field
 of a rotating black hole, as characterized by the Kerr
metric, by calculating  $P_{a,b}$ of
\r{PaPb} for all directions of wave propagation in a chosen ${\cal R}$.

For compact presentation of our results, we considered
surfaces of constant $R$ for
\begin{equation}
R_+ \leq R \leq  R_{S_+}\,. \l{ergo}
\end{equation}
The
region represented by \r{ergo} is known as  the ergosphere of the
blackhole. In regions beyond the stationary limit, i.e., $R >
R_{S_+}$, the NPV conditions \r{NPV_cond} were  found not to be
satisfied. This observation is consistent with the asymptotic
behaviour of the Kerr metric: in the limit $R \rightarrow \infty$,
Kerr spacetime converges to flat Minkowski spacetime which does
not support NPV propagation in vacuum \c{ML04a}.

\begin{figure}[!ht]
\centering \psfull \epsfig{file=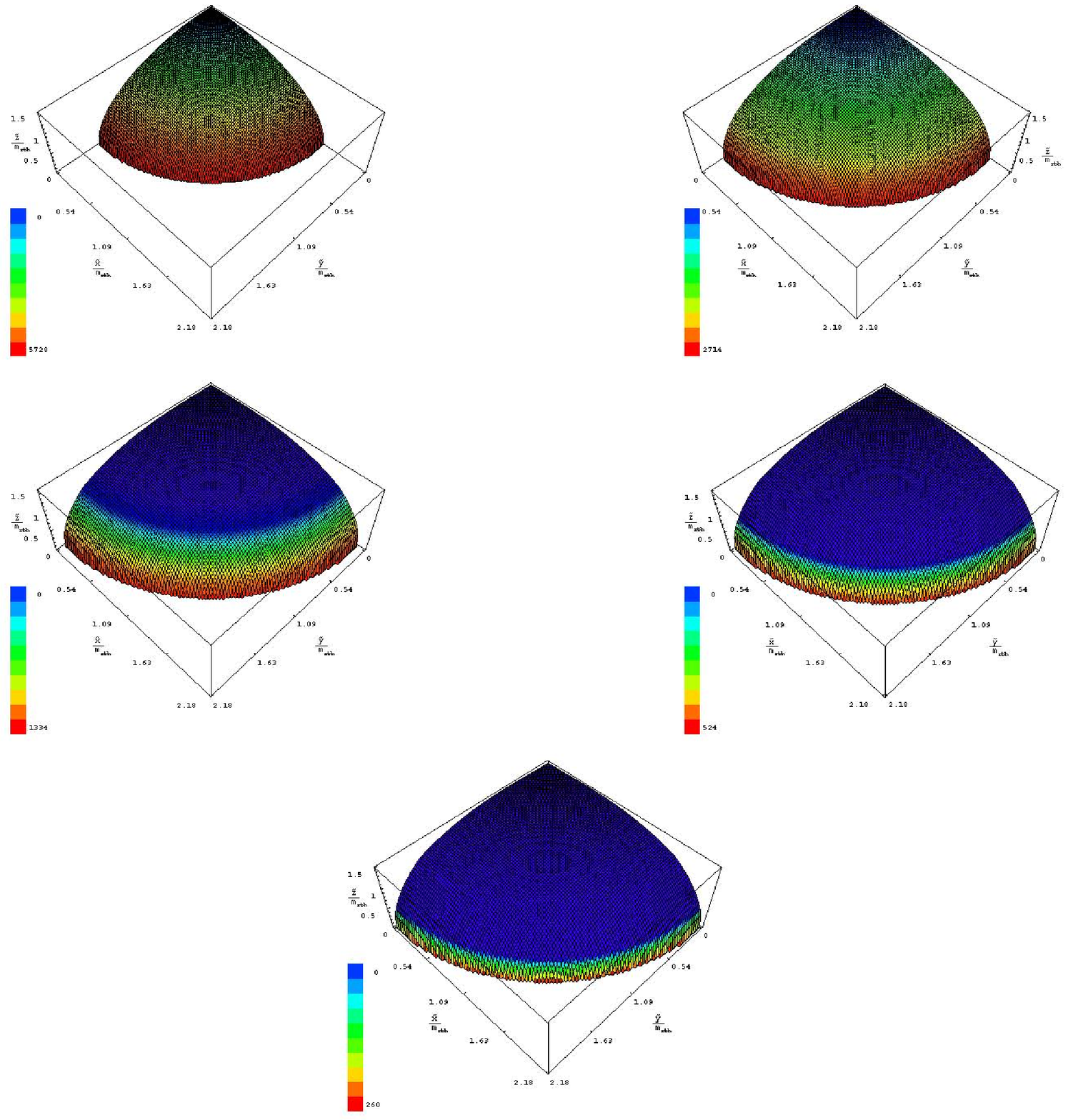,width=6.4in}
 \caption{\label{fig1} NPV maps for $\arbh = \mrbh  \sqrt{3/4}$.
See the text for an explanation of the colour mapping.
 (a) At the surface $R =  R_+$ (i.e., at the outer event
 horizon)(top left); (b) At the surface $R = 0.5 R_{+} + 0.5
 R_{S_+}$ (top right); (c) At the surface $R = 0.25 R_{+} + 0.75
 R_{S_+}$ (middle left); (d) At the surface $R = 0.1 R_{+} + 0.9
 R_{S_+}$ (middle right); and (e) At the surface $R = 0.05 R_{+} + 0.95
 R_{S_+}$ (bottom) .
  }
\end{figure}

 For the purposes of illustration, we chose  a black
hole with  $\arbh = \mrbh \sqrt{3/4}$. We explored
the feasibility of NPV propagation, in all directions denoted by
$\hat{\#k} = \le \sin \theta \cos \phi, \sin \theta \sin \phi,
\cos \theta \ri $, at $\le \tilde{x}, \tilde{y}, \tilde{z} \ri$
neighbourhoods on surfaces of constant $R$. In Figure~\ref{fig1},
the surfaces
\begin{itemize}
\item[(a)] $R = R_+$ (i.e., the outer event horizon),
\item[(b)] $R= 0.5 R_+ + 0.5 R_{S_+}$,
\item[(c)] $R= 0.25 R_+ + 0.75 R_{S_+}$,
\item[(d)] $R= 0.1 R_+ + 0.9 R_{S_+}$, and
\item[(e)] $R= 0.05 R_+ + 0.95 R_{S_+}$
\end{itemize}
are  mapped for the octant $\lec\tilde{x}
> 0,\,\tilde{y} > 0,\,\tilde{z} > 0\ric$. The three--dimensional
surface mesh is constructed from  $\tilde{z}$ heights calculated
on a 90 $\times$ 90 grid of $\le \tilde{x}, \tilde{y} \ri$ points.
At each $\le \tilde{x}, \tilde{y}, \tilde{z} \ri$ point, the
wavevector $\#k = k^\pm \, \hat{\#k}$, together with the
parameters $P_a$ and $P_b$, were computed at $1^\circ$ increments
for $ \theta \in [0^\circ ,180^\circ)$ and $ \phi \in [0^\circ
,360^\circ)$. If NPV was found to arise at a particular $( \theta,
\phi )$ orientation (as indicated by both $P_a < 0$ and $P_b < 0$)
then the  $\le \tilde{x}, \tilde{y}, \tilde{z} \ri$
 grid point was
given a `score' of one unit.\footnote{If  NPV occurs for both the
$k^+$ and $k^-$ wavenumbers for a particular $( \theta, \phi )$
orientation
 then a score of two
units was awarded.} Thus, at each $\le \tilde{x}, \tilde{y},
\tilde{z} \ri$
 grid point the maximum theoretical
score is $180 \times 360 \times 2 = 129600$. The NPV scores are
represented by colours in Figure~\ref{fig1}. We remark that the
magnitude of $\omega$ does not influence the signs of $P_{a,b}$
(nor the NPV scores).

It is clear from Figure~\ref{fig1} that NPV propagation is most
prevalent at $\le \tilde{x}, \tilde{y}, \tilde{z} \ri$
neighbourhoods close to the equator of the outer event horizon (i.e.,
at neighbourhoods  in the vicinity of the plane $\tilde{z} = 0$). The density of NPV
propagation  decreases towards the $\tilde{x} =
\tilde{y} = 0$ pole of the $R = R_+$ region. We note that  the ergosphere vanishes at  the
poles, as the outer event horizon and the
stationary limit surface coincide.

For ergospheric regions outside the outer event horizon,
distributions of NPV propagation density are mapped in
Figure~\ref{fig1}(b)--(e).  On comparing the maps in
Figure~\ref{fig1}(a)--(e),  it becomes clear that, as the constant $R$
surfaces
 approach
the stationary limit surface, the densities of NPV propagation
become progressively more concentrated around the $\tilde{z} = 0$
equator.

In order to investigate  the influence of the angular velocity
term $\arbh$ we repeated the calculations of Figure~\ref{fig1} but
with $\arbh = \mrbh /3$. The corresponding NPV propagation maps
are presented in Figure~\ref{fig2}, for the same surfaces of
constant $R$ as were used in Figure~\ref{fig1}.
  As compared with $\arbh = \mrbh
\sqrt{3/4}$, the outer event horizon and the  stationary limit
surface are closer together  for $\arbh= \mrbh /3$. The
concentration of NPV propagation at the $\tilde{z} = 0$ equator is
evidently somewhat more pronounced in the case of the thinner
ergosphere. It also noteworthy that the overall occurrence of NPV
propagation is less common for the thinner ergosphere, as
indicated by the generally lower NPV scores recorded in
Figure~\ref{fig2} as compared with Figure~\ref{fig1}.

From Figures~\ref{fig1} and \ref{fig2} it is clear that the
occurrence of NPV increases as the black hole  rotation increases.
It is therefore of interest to look at the scenario $\arbh
\rightarrow \mrbh$ for which the occurrence of NPV is maximized.
The NPV distribution maps corresponding to $\arbh = 0.99 \mrbh$
are shown in Figure~\ref{fig3}, for the same surfaces of constant
$R$ as were considered in Figures~\ref{fig1} and \ref{fig2}. In
this maximized scenario, the regions which support NPV propagation
are widely distributed throughout the ergosphere. However, we note
the relative sparsity of NPV propagation in the vicinity of the
black hole's axis of rotation.

 Extrapolating from Figures~\ref{fig1}, \ref{fig2} and \ref{fig3}, we infer that
  NPV propagation  does not occur
  \begin{itemize}
  \item[(i)] in neighbourhoods
 along the black
 hole's axis of rotation, and
 \item[(ii)] in the limit as the black hole's
  rate of rotation becomes
vanishingly small.
\end{itemize}
We note that
 the outer event
horizon and the stationary limit surface merge as the rotational
rate approaches zero, and thereby the ergosphere disappears.

 On comparing the maximum
theoretical  NPV score of 129600 with the maximum scores in
Figures~\ref{fig1},  \ref{fig2} and \ref{fig3}, it is apparent
that the conditions for  NPV propagation are satisfied for
relatively few wavevector orientations in the ergosphere. To
illustrate this point further, we considered a single
representative neighbourhood within the ergosphere, say $\tilde{x}
= 0.2 \, \mrbh$, $\tilde{y} = 1.8 \, \mrbh$ and $\tilde{z} = 0.3
\, \mrbh$, with $\arbh = \mrbh \sqrt{3/4}$ (for which $R =
1.63\,\mrbh$). We computed $P_{a,b}$ for all orientations of
$\hat{\#k}$. At orientations where $P_{a,b} < 0$ a score of one
unit was recorded. The scores are plotted against the spherical
polar coordinates $\theta$ and $\phi$ of $\hat{\#k}$ in
Figure~\ref{fig4}. Orientations at which $P_{a,b} < 0$ arises from
the $k^+$ wavenumber are coloured red, whereas those orientations
at which $P_{a,b} < 0$ occurs for the $k^-$ wavenumber are
coloured green. The $k^+$ NPV regions and the $k^-$ NPV regions
are observed in Figure~\ref{fig3} to occupy a relatively small
portion of the $ \le \theta, \phi \ri $ plane.

\section{Concluding remarks}

By using the Kerr metric in the Tamm formulation of electromagnetics in
gravitationally affected vacuum,
we have mapped
the occurrence of negative--phase velocity (NPV) planewave propagation
in the ergosphere of a rotating black hole. Our numerical results indicate that
\begin{itemize}
\item NPV propagation does not appear possible outside the stationary limit surface;
\item rotation of a black hole is required for NPV propagation;
\item NPV propagation is impossible on the axis of rotation of the black hole;
\item
regions of NPV propagation are concentrated at the equator of the
ergosphere, and are less  common towards the polar regions; and
\item an increase in   angular velocity
exaggerates the NPV concentration at the equator.
\end{itemize}
In particular, the importance of the rotational character of the black hole metric should be stressed.
In the absence of rotation the black hole metric does not support NPV propagation.
The infeasibility of NPV propagation for (i) a non--rotating black
hole; and  (ii) along the axis of rotation of a black hole, is
consistent with our earlier analytic results \c{LM04,LMS05}. In connection with this issue,
we note that NPV propagation can however arise in  Schwarzschild--de Sitter spacetime \c{Schwarzs}.

Let us comment upon the  applicability of the NPV condition
\r{NPV_cond}. Suppose that the neighbourhood ${\cal R}$ has
representative linear dimensions given by $\rho$. In order for the
nonuniform metric $g_{\alpha \beta}$ to be approximated by the
uniform metric $\tilde{g}_{\alpha \beta}$ throughout ${\cal R}$,
it is necessary  that $\rho$ be small compared to the radius of
curvature of the $g_{\alpha \beta}$ spacetime. A convenient
measure of the inverse  radius of spacetime curvature squared is
provided by the nonzero components of the Riemann tensor. These
components are of the order of $R^{-2}$ for the Kerr metric
\c{Chandra}. Therefore, we have $\rho \ll R$. Since the
neighbourhood ${\cal R}$ is also required to be large compared
with
 electromagnetic wavelengths, as given by $2 \pi/ |k|$,
we see that \r{NPV_cond}  holds in the regime
\begin{equation} \l{applicability}
\frac{2 \pi}{ | k | } \ll  R.
\end{equation}
For black holes of the mass of our sun, the condition
\r{applicability} translates to wavelengths of the order of 100
metres or less, whereas for supermassive black  holes of the type
believed to lie at the centre of galaxies the NPV condition
\r{NPV_cond} holds for wavelengths of the order of $10^7$
kilometres or less, typically \c{supermassive}.

The NPV phenomenon reported in the preceding sections concerns the
propagation of electromagnetic plane waves in the ergosphere. The
issue of possible generalizations to scalar and/or gravitational
waves is a matter for future study. However, we note in this
context that the
 Teukolsky equation yields general (nonplanar)
 solutions which support negative wavenumbers for
scalar, electromagnetic and gravitational waves
 in the ergosphere of a Kerr black hole \c{Teukolsky1}--\c{Teukolsky3}.

It is  of interest to speculate whether the NPV propagation
demonstrated here may arise for so--called acoustic black holes
\c{Barcelo}, typically studied in the cylindrical
geometry \c{Visser}. While cylindrical acoustic black holes can support superradiant
scattering analogously to the equatorial slice of the Kerr black hole
\c{Visser,Berti}, the propagation of NPV acoustic waves  for  acoustic
black holes has yet to be established.

We close by emphasizing that the description of NPV
presented herein arises from the classical treatment of
electromagnetic waves in curved spacetime, within the short
wavelength regime indicated by \r{applicability}. This NPV
phenomenon is distinct from superradiant scattering
in the ergosphere at long wavelengths \c{SML_05}.

\bigskip

\noindent{\bf Acknowledgement}

\noindent
SS acknowledges EPSRC for support under grant GR/S60631/01. We thank
two anonymous referees for comments that helped us improve this communication.

\newpage

\vspace{5mm}

\newpage

 \setcounter{figure}{1}

\begin{figure}[!ht]
\centering \psfull \epsfig{file=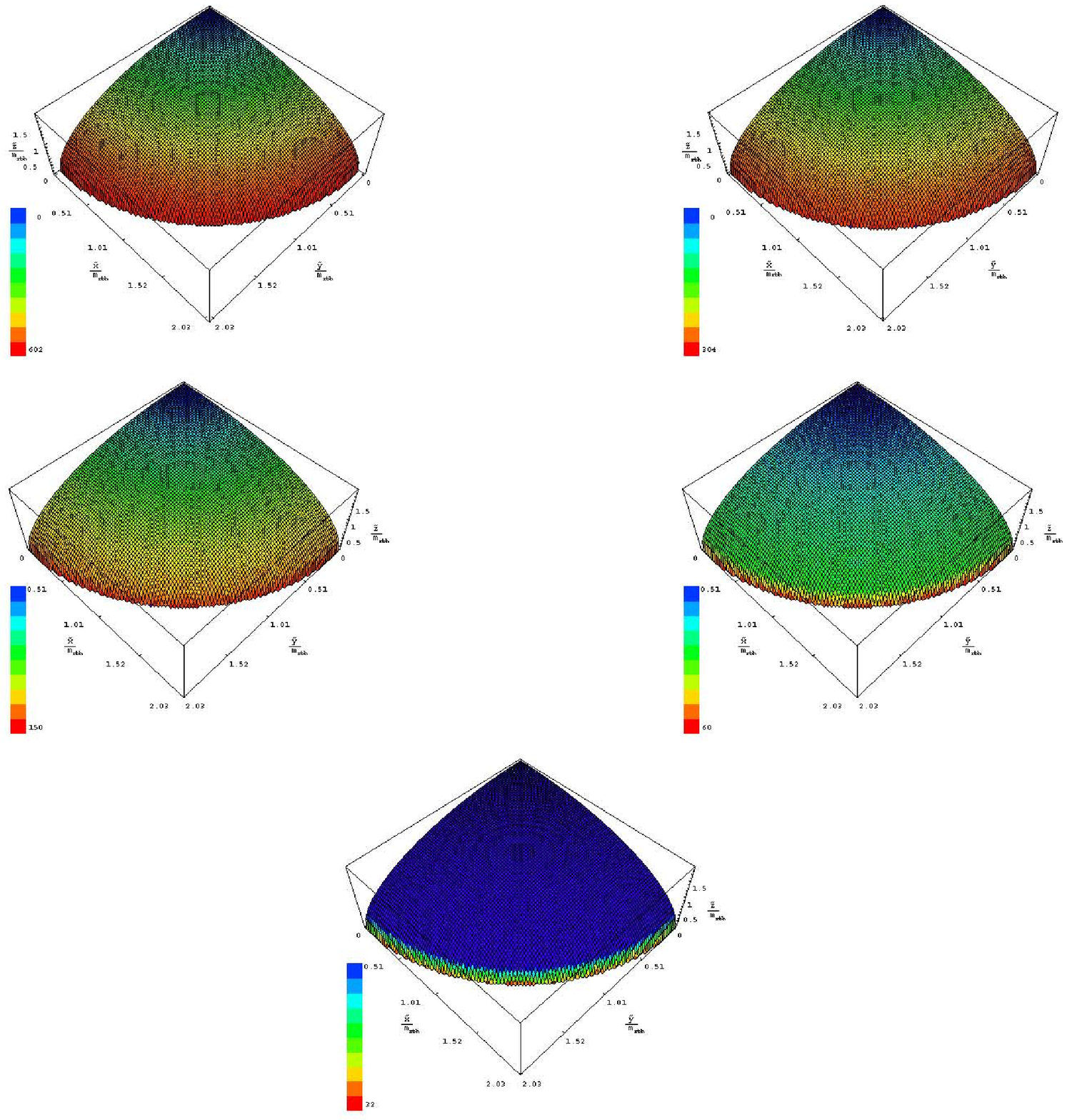,width=6.4in}
 \caption{\label{fig2}
NPV maps for $\arbh = \mrbh /3$. See the text for an explanation
of the colour mapping.
 (a) At the surface $R =  R_+$ (i.e., at the outer event
 horizon) (top left); (b) At the surface $R = 0.5 R_{+} + 0.5
 R_{S_+}$ (top right); (c) At the surface $R = 0.25 R_{+} + 0.75
 R_{S_+}$ (middle left); (d) At the surface  $R = 0.1 R_{+} +
 0.9
 R_{S_+}$ (middle right); and (e) At the surface $R = 0.05 R_{+} + 0.95
 R_{S_+}$ (bottom).
  }
\end{figure}

\newpage

 \setcounter{figure}{2}

\begin{figure}[!ht]
\centering \psfull \epsfig{file=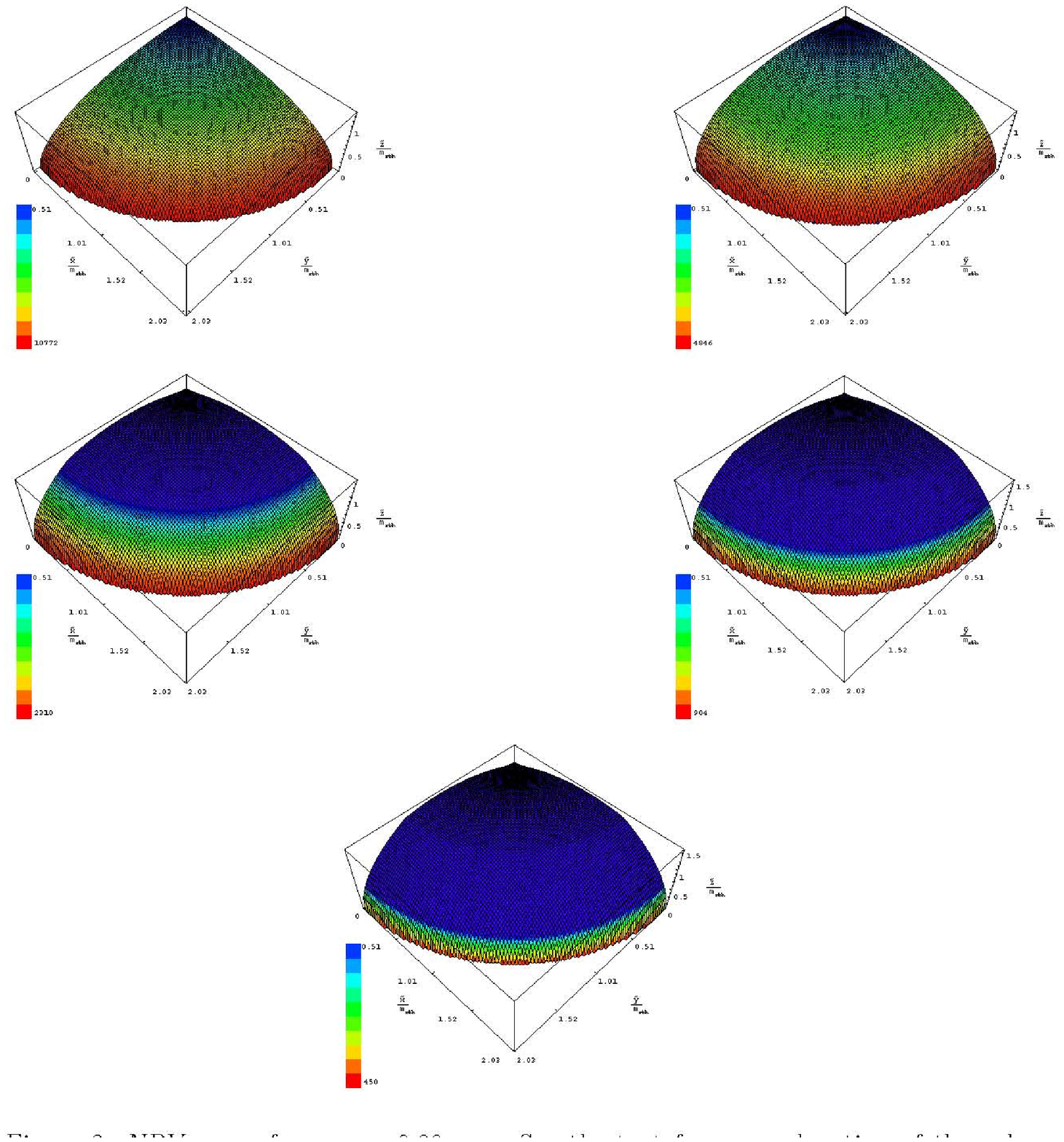,width=6.4in}
 \caption{\label{fig3}
NPV maps for $\arbh = 0.99 \mrbh $. See the text for an
explanation of the colour mapping.
 (a) At the surface $R =  R_+$ (i.e., at the outer event
 horizon) (top left); (b) At the surface $R = 0.5 R_{+} + 0.5
 R_{S_+}$ (top right); (c) At the surface $R = 0.25 R_{+} + 0.75
 R_{S_+}$ (middle left); (d) At the surface  $R = 0.1 R_{+} +
 0.9
 R_{S_+}$ (middle right); and (e) At the surface $R = 0.05 R_{+} + 0.95
 R_{S_+}$ (bottom).
  }
\end{figure}

\newpage

 \setcounter{figure}{3}
\begin{figure}[!ht]
\centering \psfull \epsfig{file=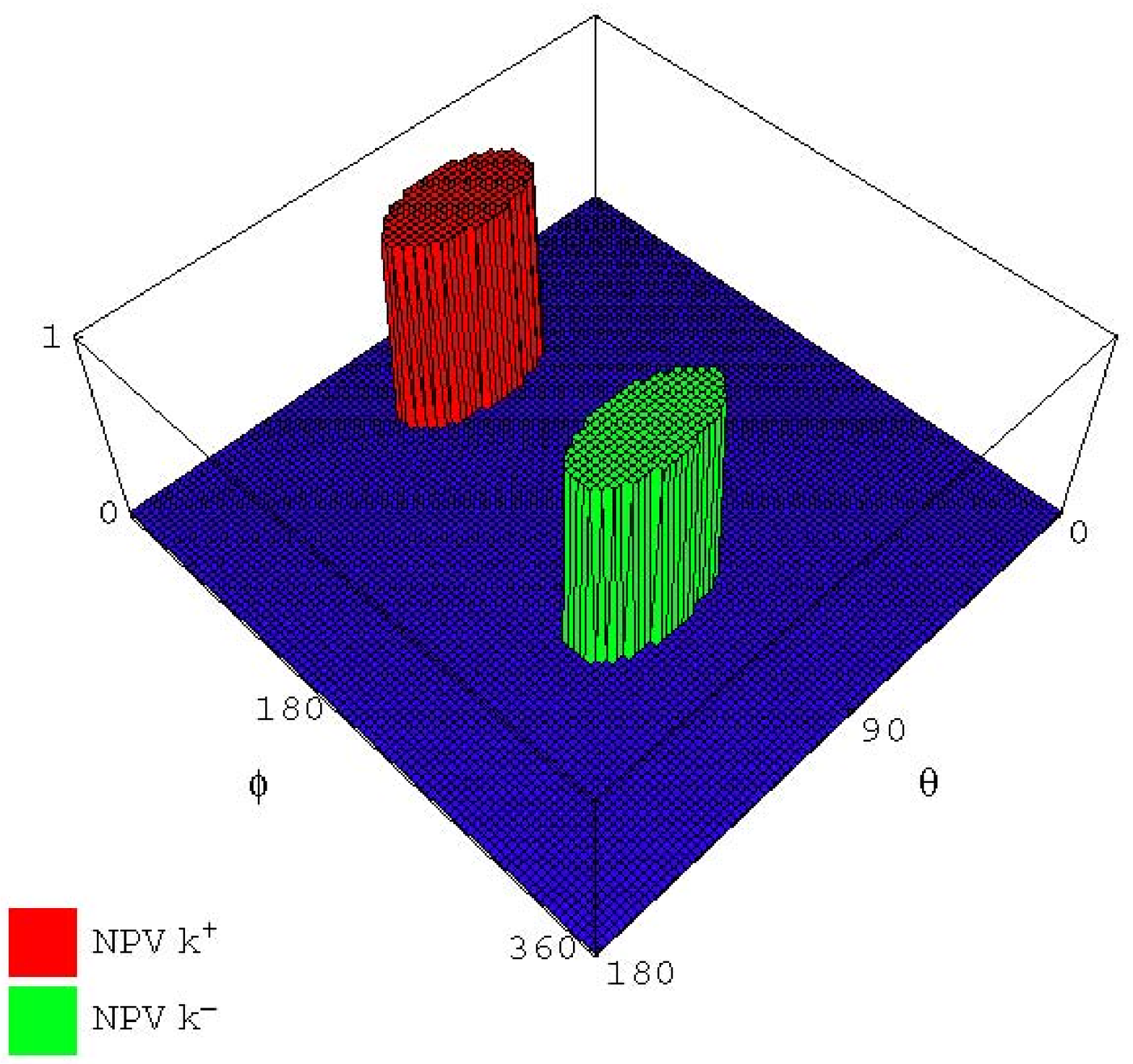,width=3.0in}
 \caption{\label{fig4} NPV scores plotted as heights, against the $\hat{\#k}$
 orientational angles $\theta$ and $\phi$, in the neighbourhood of
$(\tilde{x} = 0.2 \, \mrbh,\,\tilde{y} = 1.8 \, \mrbh,\,
\tilde{z} = 0.3 \, \mrbh )$, with $\arbh = \mrbh  \sqrt{3/4}$. NPV
arising from the $k^+$ wavenumber is coded red; NPV arising
from the $k^-$ wavenumber is coded green.
  }
\end{figure}


\begin{thebibliography}{99}

\bibitem{LMW02}
Lakhtakia A, McCall M W, and  Weiglhofer W S 2002
Brief overview of recent developments on negative phase--velocity
mediums (alias left--handed materials)
\emph{AE\"U Int. J. Electron. Commun.} {\bf 56} 407--410

\bibitem{LMW03}
Lakhtakia A, McCall M W, and  Weiglhofer W S 2003 Negative
phase--velocity mediums
  \emph{Introduction to Complex Mediums for Optics and
Electromagnetics}  ed  WS Weiglhofer and A Lakhtakia (Bellingham,
WA, USA: SPIE  Press) pp347--363

\bibitem{Pendry04}
Pendry J B  2004 Negative
refraction \emph{Contemp. Phys.} {\bf 45} 191--202


\bibitem{SSS}
Shelby R A,  Smith D R and Schultz S 2001 Experimental
verification of a negative index of refraction \emph{Science} {\bf
292} 77--79

\bibitem{LM04}
Lakhtakia A and Mackay T G 2004 Towards gravitationally assisted
negative refraction of light by vacuum
\emph{J. Phys. A: Math. Gen.} {\bf 37} L505--L510; corrigendum 2004
{\bf 37} 12093


\bibitem{LMS05}
Lakhtakia A, Mackay  T G and  Setiawan S 2005
Global and local perspectives of gravitationally assisted negative--phase--velocity
propagation of electromagnetic waves in vacuum
\emph{Phys. Lett. A}
{\bf 336} 89--96


\bibitem{LL}
Landau L D and Lifshitz E M 1975 {\em The Classical Theory of Fields}  (Oxford, UK:
Claredon Press) \S90


\bibitem{Skrotskii}
Skrotskii  G V 1957 The influence of gravitation on the propagation of light
\emph{Soviet Phys.--Dokl.\/} {\bf 2}  226--229


\bibitem{Plebanski}
Plebanski J 1960 Electromagnetic waves in gravitational fields
\emph{ Phys. Rev.\/} {\bf 118} 1396--1408

\bibitem{Moller}
M\o ller C 1972 {\em The Theory of Relativity} 2nd ed. (Oxford,
UK: Claredon Press)

\bibitem{Mashhoon}
Mashhoon B 1973 Scattering of electromagnetic radiation from a black hole
\emph{Phys. Rev. D\/} {\bf 7} 2807--2814


 \bibitem{SS}
Schleich W  and Scully M O 1984 General relativity and modern optics
 \emph{New Trends in Atomic Physics}
ed  G Grynberg  and R Stora  (Amsterdam, Holland:
Elsevier Science Publishers) pp995--1124

\bibitem{Burko}
Burko  L M 2002 Self-interaction near dielectrics \emph{Phys. Rev.
E\/} {\bf 65} 046618


\bibitem{MLS05}
Mackay  T G, Lakhtakia  A  and  Setiawan S 2005
Gravitation and electromagnetic wave propagation with negative phase velocity
 \emph{New J. Phys.} {\bf 7} 75

\bibitem{Kerr}
Kerr R P 1963 Gravitational field of a spinning mass as an example
of algebraically special metrics \emph{Phys. Rev. Lett.} {\bf 11}
237--238

\bibitem{Chen}
Chen H C 1983 {\em Theory of Electromagnetic Waves\/} (New York,
USA: McGraw--Hill)

\bibitem{Inverno}
d'Inverno R 1992 {\em Introducing Einstein's Relativity\/}
(Oxford, UK: Clarendon Press) Chap 19


\bibitem{Lopt92}
Lakhtakia A  1992 General schema for the Brewster conditions
\emph{Optik} {\bf 90}  184--186


\bibitem{ML04a}
Mackay T G and Lakhtakia A 2004 Negative phase velocity in a
uniformly moving, homogeneous, isotropic, dielectric--magnetic
medium \emph{J. Phys. A: Math. Gen.} {\bf 37}  5697--5711



\bibitem{Schwarzs}
Mackay T G, Lakhtakia A and Setiawan S
 2005 Electromagnetic waves with negative phase velocity in Schwarzschild-de Sitter spacetime
\emph{Europhys. Lett.} (accepted for publication)

\bibitem{Chandra}
Chandrasekhar S  1983 {\em The Mathematical Theory of Black
Holes\/} (Oxford, UK: Claredon Press)

\bibitem{supermassive}
New K C B and Shapiro S L 2001 The formation of supermassive black holes
and the evolution of supermassive stars
\emph{Class. Quantum Grav.} {\bf 18} 3965--3975

\bibitem{Teukolsky1}
Teukolsky S A 1973
Perturbations of a rotating black hole. I Fundamental
equations for gravitational electromagnetic, and neutrino--field perturbations
\emph{Astrophys. J.} {\bf 185}  635--647

\bibitem{Teukolsky2}
Press W H and
Teukolsky S A 1973
Perturbations of a rotating black hole. II Dynamical stability of the Kerr metric
\emph{Astrophys. J.} {\bf 185}  649--673

\bibitem{Teukolsky3}
Teukolsky S A and Press W H  1974
Perturbations of a rotating black hole. III Interaction of the hole with gravitational and elecromagnetic
radiation
\emph{Astrophys. J.} {\bf 193}  443--461

\bibitem{Barcelo}
Barcel\'o C, Liberati S, Sonego S and Visser M 2004 Causal
structure of analogue spacetimes \emph{New J. Phys.} {\bf 6} 186

\bibitem{Visser}
Visser M 1998
Acoustic black holes: horizons, ergopsheres and Hawking radiation
\emph{Class. Quant. Grav.} {\bf 15}  1767--1791

\bibitem{Berti}
Berti E 2005
Black holes in a bath tub
\emph{J. Phys.: Conf. Series} {\bf 8} 101--105


\bibitem{SML_05} Setiawan S, Mackay T G and Lakhtakia A
 2005 A comparison of superradiance
 and negative--phase--velocity
 phenomenons in the ergosphere
 of a rotating black hole
  \emph{Phys. Lett. A} {\bf 341} 15--21


\end{thebibliography}
\end{document}